\documentclass{XrU2005}
\usepackage{graphicx}
\title{XMM-Newton and Deep Optical Observations of the OTELO fields: 
the Groth-Westphal Strip}

\author[1]{Miguel S\'anchez-Portal}
\author[2]{Ana M. P\'erez-Garc\'\i a}
\author[2,3]{Jordi Cepa-Nogu\'e}
\author[4]{Emilio Alfaro}
\author[2]{H\'e ctor Casta\~neda }
\author[5]{Jes\'us Gallego}
\author[6]{J. J. Gonz\'alez-Gonz\'alez}
\author[7]{J. Ignacio Gonz\'alez-Serrano}
\affil[1]{Herschel Science Centre - INSA/ESAC, P.O. Box 50727, Madrid 28080, Spain}
\affil[2]{Instituto de Astrof\'\i sica de Canarias, E-38205, La Laguna, Tenerife, Spain }
\affil[3]{Universidad de La Laguna, La Laguna, Tenerife, Spain }
\affil[4]{Instituto de Astrof\'\i sica de Andaluc\'\i a (CSIC), Granada, Spain }
\affil[5]{Departamento de Astrof\'\i sica, Universidad Complutense de Madrid, Spain }
\affil[6]{Instituto de Astronom\'\i a, UNAM, M\'exico DF, M\'exico }
\affil[7]{Instituto de F\'\i sica de Cantabria, CSIC-UNICAN, Santander, Spain}

\begin{document}


\maketitle

\begin{abstract}
We present a preliminary analysis of public EPIC data of one of the OTELO targets,
the Groth-Westphal strip, gathered from the XMM-Newton Science Archive (XSA). EPIC
images are combined with optical $BVRI$ data from our broadband survey carried 
out with the 4.2m WHT at La Palma.
\end{abstract}

\section{The OTELO Project and its Science}

OTELO \citep{cepa} will search for emission line objects using OSIRIS 
tunable filters 
at the 10m GTC telescope in La Palma in selected atmospheric windows (centred at 
the H$_{\alpha}$ line at z = 0.24 and z = 0.4) that 
are relatively free of sky emission lines. A total area of more than one square 
degree will be observed. The survey technique will allow for separation of the
H$_{\alpha}$ and [N{\small II}] lines and therefore 
AGNs from starburst galaxies.
A 5$\sigma$ depth of $8 \times 10^{-18}$ erg cm$^{-2}$ s$^{-1}$ will make 
OTELO the deepest emission line survey to date.
OTELO science includes  the evolution 
of galaxies, the evolution of star formation in the Universe, chemical evolution 
of galaxies, QSO spatial density determination, AGN evolution, the low-end of the 
galaxy luminosity function, galactic Astronomy etc.
A complementary optical broadband survey 
is currently on-going (morphology, photometrical redshifts). 
It is intended to complement the optical survey with other ground or space-based
facilities, from X-rays (XMM-Newton, Chandra) to FIR and sub-mm (Herschel, GTM). The
present work is being performed in the framework of the multiwavelength study of the
OTELO fields.

\section{Observations and data reduction}

XMM-Newton observations of the Groth-Westphal strip 
\citep{groth} were collected from the XMM Science 
Archive (XSA). The EPIC observations were reprocessed 
using SAS v6.0.0 {\tt emproc} and {\tt epproc} 
standard procedures. High-radiation intervals were removed by inspecting the 
count rate curves. GTI files were created by means of the {\tt tabgtigen} SAS task. 
These files were further included in standard filtering expressions.
The observations were co-added by means of the SAS {\tt merge} procedure. 
Attitude files were also merged. Final exposure times were about 82~ksec and 
70~ksec in MOS and PN, respectively. Three energy bands were defined:
\mbox{0.5 - 2 keV} (soft band), \mbox{2 - 4.5 keV} (medium band) and 
\mbox{4.5 - 10 keV} (hard band). Sources were detected by means of 
the {\tt edetect\_chain} SAS procedure. We imposed a likehood parameter 
$ML = -\ln (1-P) > 14$, were $P$ is the probability that the source exists.
Furthermore, we considered only those sources for which $flux/err_{flux} > 2$, 
lying within the optical FOV. 
We have detected 75 sources fulfilling these conditions. 
Two hardness ratios have been also computed by this SAS procedure.

Optical $BVRI$ observations were carried out at the prime focus of 
the 4.2m William Herschel Telescope (WHT) 
of the Observatorio del Roque de los Muchachos (La Palma). 
Image size is \mbox{16' $\times$ 16'} with a plate scale of 0.24 arcsec/pixel. 
Total exposure time at each of the three pointing directions is 9000~sec per filter.
Reduction process followed standard steps using IRAF packages.
Photometric calibration was obtained with several Landolt standard fields.
Absolute astrometry was performed using the USNO B1 catalogue.
Sources were extracted using Sextractor 2.2 \citep{bertin}.
The 50\% detection level is 25.3 in B, 25.3 in V, 24.7 in R and 23.5 in I.
Details on the optical observations can be found in \citet{ana}.

\section{Source Detection}

We have matched X-rays and optical sources by searching in 
\mbox{6" $\times$ 6"} boxes centred in the X-rays sources coordinates. 
Upon comparison  with published Chandra coordinates 
\citep{nandra} we found a bulk shift of $\langle \Delta \alpha \rangle = 4.5"$ and 
 $\langle \Delta \delta \rangle = -2.7"$. 
After correcting for this global image shift, we found: 
{\em (a)} Unique match in all photometric bands	for 43	sources (57.3\%);
 {\em (b)} partial match (not detected in all bands) for 10 sources (13.3\%); 
  {\em (c)} multiple match for 10 sources (13.3\%); 
  {\em (d)} partial match + multiple match for 2 sources (2.7\%); and
 {\em (e)} No optical counterpart found for 10 sources (13.3\%)

\section{Results}

Since photometric redshift values are still not available, we have concentrated our
analysis in distance-independent parameters, as described below.

\subsection{X/O Ratio Analysis}

\begin{figure}
\centering
\includegraphics[angle=270,width=1.1\linewidth]{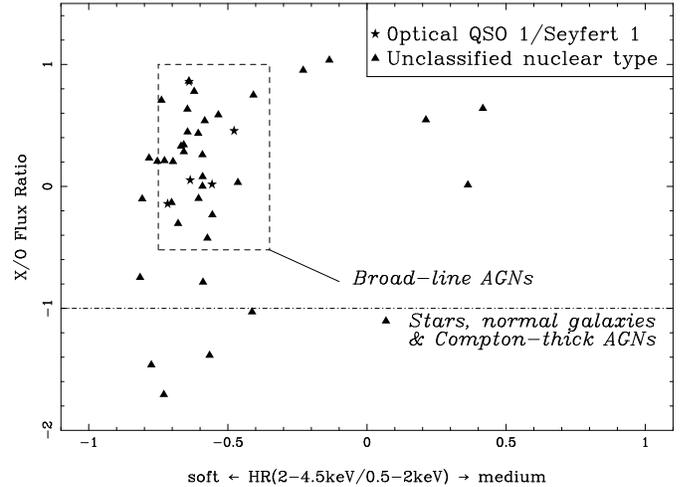}
\caption{X/O relation vs. hardness ratio}
\label{fig1}
\end{figure}

The X-rays to optical flux ratio (X/O) is a powerful means to discriminate
between different classes of X-rays sources 
(Fiore et~al., 2003; Della Ceca et~al., 2004). 
Figure \ref{fig1} shows a diagnostic diagram combining the 0.5-4.5~keV to optical
R-band flux ratio and one of the
hardness ratios. The dashed-dotted line corresponds to X/O = 0.1, typical of
coronal-emitting stars, normal galaxies and heavily absorbed AGNs. Five of our sources 
(11.6\%) lie 
below this line, and therefore are likely either normal galaxies or Compton-thick AGNs. 
The dashed-line box
corresponds to the region containing the 85\% of the optically identified 
broad-line AGNs in the XMM-Newton Bright Source Sample (BSS) \citep{dellaceca}. 29 of our
sources (67.4\%) fall within the broad-line AGN region defined by this box. Objects with
harder HR and large X/O (i.e. those to the right of the broad-line box) 
are likely narrow-line AGNs, according to the BSS and HBSS diagnostics from \citet{dellaceca}.

\subsection{B/T Relation}

\begin{figure}
\centering
\includegraphics[angle=270,width=0.8\linewidth]{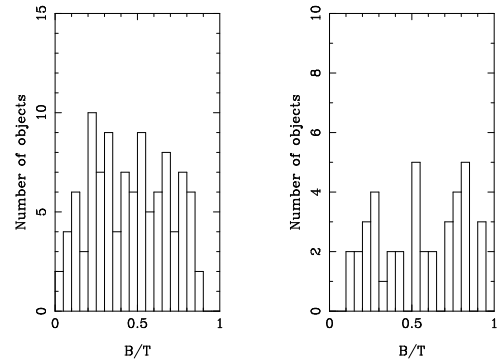}
\caption{B/T relation in the local Universe (left) and in the X-rays selected sample}
\label{fig2}
\end{figure}

We have derived the B/T (bulge-to-total luminosity) relation for the sample objects,
comparing its distribution with a sample of active galaxies 
in the local Universe \citep{yo}, as shown in figure \ref{fig2}. 
While not incompatible, B/T distributions are likely different, 
as proven by means of a K-S test: null hypothesis 
(that both samples are drawn from the same distribution) significance 
is only 17\%.  If we exclude LINERs from the local Universe sample (i.e.
considering only Seyfert galaxies), the null hypothesis 
significance is higher but still reduced, 39\%. 
The X-rays selected sample tends to higher B/T values and lacks very 
low B/T objects (generally present in latest Hubble types, T $\ge$ 4). 
Finally, we don't find any correlation between X-rays hardness ratios and the 
B/T ratio. 

\end{document}